\documentclass[twocolumn]{aastex631}
\usepackage{amsmath}
\usepackage{graphics}
\usepackage{graphicx}
\usepackage{amssymb}

\begin{document}

\title{Possible First Detection of Gyroresonance Emission from a Coronal Mass Ejection in the Middle Corona}
\correspondingauthor{Surajit Mondal}
\email{surajit.mondal@njit.edu}

\author[0000-0002-2325-5298]{Surajit Mondal}
\affiliation{Center for Solar-Terrestrial Research, New Jersey Institute of Technology, \\
323 M L King Jr Boulevard, Newark, NJ 07102-1982, USA}

\author[0000-0002-0660-3350]{Bin Chen}
\affiliation{Center for Solar-Terrestrial Research, New Jersey Institute of Technology, \\
323 M L King Jr Boulevard, Newark, NJ 07102-1982, USA}

\author[0000-0002-1810-6706]{Xingyao Chen}
\affiliation{Center for Solar-Terrestrial Research, New Jersey Institute of Technology, \\
323 M L King Jr Boulevard, Newark, NJ 07102-1982, USA}

\author[0000-0003-2872-2614]{Sijie Yu}
\affiliation{Center for Solar-Terrestrial Research, New Jersey Institute of Technology, \\
323 M L King Jr Boulevard, Newark, NJ 07102-1982, USA}

\author{Dale Gary}
\affiliation{Center for Solar-Terrestrial Research, New Jersey Institute of Technology, \\
323 M L King Jr Boulevard, Newark, NJ 07102-1982, USA}

\author[0000-0001-6855-5799]{Peijin Zhang}
\affiliation{Center for Solar-Terrestrial Research, New Jersey Institute of Technology, \\
323 M L King Jr Boulevard, Newark, NJ 07102-1982, USA}
\affiliation{Cooperative Programs for the Advancement of Earth System Science, University Corporation for Atmospheric Research, Boulder, CO, USA}

\author{Marin M. Anderson}
\affiliation{Owens Valley Radio Observatory, California Institute of Technology, Big Pine, CA 93513, USA}
\affiliation{Jet Propulsion Laboratory, California Institute of Technology, Pasadena, CA 91011, USA}
\author{Judd D. Bowman}
\affiliation{School of Earth and Space Exploration, Arizona State University, Tempe, AZ 85287, USA}
\author{Ruby Byrne}
\affiliation{Cahill Center for Astronomy and Astrophysics, California Institute of Technology, Pasadena, CA 91125, USA}
\affiliation{Owens Valley Radio Observatory, California Institute of Technology, Big Pine, CA 93513, USA}
\author{Morgan Catha}
\affiliation{Owens Valley Radio Observatory, California Institute of Technology, Big Pine, CA 93513, USA}
\author{Sherry Chhabra}
\affiliation{Center for Solar-Terrestrial Research, New Jersey Institute of Technology, \\
323 M L King Jr Boulevard, Newark, NJ 07102-1982, USA}
\affiliation{George Mason University, Fairfax, VA 22030, USA}
\author{Larry D'Addario}
\affiliation{Cahill Center for Astronomy and Astrophysics, California Institute of Technology, Pasadena, CA 91125, USA}
\affiliation{Owens Valley Radio Observatory, California Institute of Technology, Big Pine, CA 93513, USA}
\author{Ivey Davis}
\affiliation{Cahill Center for Astronomy and Astrophysics, California Institute of Technology, Pasadena, CA 91125, USA}
\affiliation{Owens Valley Radio Observatory, California Institute of Technology, Big Pine, CA 93513, USA}
\author{Jayce Dowell}
\affiliation{University of New Mexico, Albuquerque, NM 87131, USA}
\author{Katherine Elder}
\affiliation{School of Earth and Space Exploration, Arizona State University, Tempe, AZ 85287, USA}
\author{Gregg Hallinan}
\affiliation{Cahill Center for Astronomy and Astrophysics, California Institute of Technology, Pasadena, CA 91125, USA}
\affiliation{Owens Valley Radio Observatory, California Institute of Technology, Big Pine, CA 93513, USA}
\author{Charlie Harnach}
\affiliation{Owens Valley Radio Observatory, California Institute of Technology, Big Pine, CA 93513, USA}
\author{Greg Hellbourg}
\affiliation{Cahill Center for Astronomy and Astrophysics, California Institute of Technology, Pasadena, CA 91125, USA}
\affiliation{Owens Valley Radio Observatory, California Institute of Technology, Big Pine, CA 93513, USA}
\author{Jack Hickish}
\affiliation{Real-Time Radio Systems Ltd, Bournemouth, Dorset BH6 3LU, UK}
\author{Rick Hobbs}
\affiliation{Owens Valley Radio Observatory, California Institute of Technology, Big Pine, CA 93513, USA}
\author{David Hodge}
\affiliation{Cahill Center for Astronomy and Astrophysics, California Institute of Technology, Pasadena, CA 91125, USA}
\author{Mark Hodges}
\affiliation{Owens Valley Radio Observatory, California Institute of Technology, Big Pine, CA 93513, USA}
\author{Yuping Huang}
\affiliation{Cahill Center for Astronomy and Astrophysics, California Institute of Technology, Pasadena, CA 91125, USA}
\affiliation{Owens Valley Radio Observatory, California Institute of Technology, Big Pine, CA 93513, USA}
\author{Andrea Isella}
\affiliation{Department of Physics and Astronomy, Rice University, Houston, TX 77005, USA}
\author{Daniel C. Jacobs}
\affiliation{School of Earth and Space Exploration, Arizona State University, Tempe, AZ 85287, USA}
\author{Ghislain Kemby}
\affiliation{Owens Valley Radio Observatory, California Institute of Technology, Big Pine, CA 93513, USA}
\author{John T. Klinefelter}
\affiliation{Owens Valley Radio Observatory, California Institute of Technology, Big Pine, CA 93513, USA}
\author{Matthew Kolopanis}
\affiliation{School of Earth and Space Exploration, Arizona State University, Tempe, AZ 85287, USA}
\author{Nikita Kosogorov}
\affiliation{Cahill Center for Astronomy and Astrophysics, California Institute of Technology, Pasadena, CA 91125, USA}
\affiliation{Owens Valley Radio Observatory, California Institute of Technology, Big Pine, CA 93513, USA}
\author{James Lamb}
\affiliation{Owens Valley Radio Observatory, California Institute of Technology, Big Pine, CA 93513, USA}
\author{Casey Law}
\affiliation{Cahill Center for Astronomy and Astrophysics, California Institute of Technology, Pasadena, CA 91125, USA}
\affiliation{Owens Valley Radio Observatory, California Institute of Technology, Big Pine, CA 93513, USA}
\author{Nivedita Mahesh}
\affiliation{Cahill Center for Astronomy and Astrophysics, California Institute of Technology, Pasadena, CA 91125, USA}
\affiliation{Owens Valley Radio Observatory, California Institute of Technology, Big Pine, CA 93513, USA}
\author{Brian O'Donnell}
\affiliation{Center for Solar-Terrestrial Research, New Jersey Institute of Technology, \\
323 M L King Jr Boulevard, Newark, NJ 07102-1982, USA}
\author{Kathryn Plant}
\affiliation{Owens Valley Radio Observatory, California Institute of Technology, Big Pine, CA 93513, USA}
\affiliation{Jet Propulsion Laboratory, California Institute of Technology, Pasadena, CA 91011, USA}
\author{Corey Posner}
\affiliation{Owens Valley Radio Observatory, California Institute of Technology, Big Pine, CA 93513, USA}
\author{Travis Powell}
\affiliation{Owens Valley Radio Observatory, California Institute of Technology, Big Pine, CA 93513, USA}
\author{Vinand Prayag}
\affiliation{Owens Valley Radio Observatory, California Institute of Technology, Big Pine, CA 93513, USA}
\author{Andres Rizo}
\affiliation{Owens Valley Radio Observatory, California Institute of Technology, Big Pine, CA 93513, USA}
\author{Andrew Romero-Wolf}
\affiliation{Jet Propulsion Laboratory, California Institute of Technology, Pasadena, CA 91011, USA}
\author{Jun Shi}
\affiliation{Cahill Center for Astronomy and Astrophysics, California Institute of Technology, Pasadena, CA 91125, USA}
\author{Greg Taylor}
\affiliation{University of New Mexico, Albuquerque, NM 87131, USA}
\author{Jordan Trim}
\affiliation{Owens Valley Radio Observatory, California Institute of Technology, Big Pine, CA 93513, USA}
\author{Mike Virgin}
\affiliation{Owens Valley Radio Observatory, California Institute of Technology, Big Pine, CA 93513, USA}
\author{Akshatha Vydula}
\affiliation{School of Earth and Space Exploration, Arizona State University, Tempe, AZ 85287, USA}
\author{Sandy Weinreb}
\affiliation{Cahill Center for Astronomy and Astrophysics, California Institute of Technology, Pasadena, CA 91125, USA}
\author{Scott White}
\affiliation{Owens Valley Radio Observatory, California Institute of Technology, Big Pine, CA 93513, USA}
\author{David Woody}
\affiliation{Owens Valley Radio Observatory, California Institute of Technology, Big Pine, CA 93513, USA}
\author{Thomas Zentmeyer}
\affiliation{Owens Valley Radio Observatory, California Institute of Technology, Big Pine, CA 93513, USA}

\begin{abstract}

{Routine measurements of the magnetic field of coronal mass ejections (CMEs) have been a key challenge in solar physics. Making such measurements is important both from a space weather perspective and for understanding the detailed evolution of the CME. In spite of significant efforts and multiple proposed methods, achieving this goal has not been possible to date. Here we report the first possible detection of gyroresonance emission from a CME. Assuming that the emission is happening at the third harmonic, we estimate that the magnetic field strength ranges from 7.9--5.6 G between 4.9-7.5 $R_\odot$. We also demonstrate that this high magnetic field is not the average magnetic field inside the CME, but most probably is related to small magnetic islands, which are also being observed more frequently with the availability of high-resolution and high-quality white-light images.}

\end{abstract}


\section{Introduction}

Coronal mass ejections (CMEs) represent the most energetic explosions in the Solar System and are characterized by massive eruptions of magnetized plasma from the solar atmosphere into the interplanetary space. CMEs travel through the interplanetary space and occasionally reach Earth. Due to their high momentum and magnetic field, CMEs are the primary drivers of space weather events in the near-Earth environment. A primary determinant of the geo-effectiveness of a CME is its magnetic field. However, at present, the CME magnetic field can only be measured in a routine manner using in situ data. Barring the chance encounters with satellites in the interplanetary space, like Solar Orbiter and Parker Solar Probe, almost all such measurements are made when the CME is very close to Earth. However, to understand the detailed evolution of the CME through the interplanetary space, as well as to provide early warning of space-weather effects, it is important to measure the CME magnetic field close to the Sun. Hence, the development of suitable remote-sensing techniques is critical.

Multiple techniques have been proposed in the literature for measuring the CME magnetic field when the CME is close to the Sun. Some of the popular techniques for determining the average magnetic field use for example the band splitting of type II bursts observed in the solar radio dynamic spectrum \citep[e.g.,][]{smerd75,gary1984,cunha-silva15,kumari17a,mahrous18}, circular polarization of moving type IV bursts \citep{raja14}, and the standoff distance from extreme ultraviolet and optical images \citep{gopalswamy11,gopalswamy12,poomvises12}. However, these techniques can only estimate the spatially averaged magnetic field, with the extent of spatial averaging varying between methods.
\citet{susino15} developed and successfully applied a technique for estimating the spatially varying magnetic field at the shock front under plausible assumptions. 

Gyrosynchrotron emission from CMEs is a very promising probe for measuring the spatially varying magnetic field entrained in the CME \citep{bastian2001, maia2007, tun2013, bain2014, mondal2020, kansabanik2023, chhabra2021}. It can also be used to determine the nonthermal electron population associated with the CME. Faraday rotation due to the CME material along the line of sight when observing linearly polarized emission from celestial sources or artificial satellites has also been used to estimate the CME magnetic field \citep[see][for a review]{kooi2022}. 


Here we present the first possible detection of thermal gyroresonance emission from a CME in the middle corona, which may serve as a new method for measuring the CME magnetic field. Gyroresonance emission is produced by thermal electrons gyrating in the magnetic field. The spectrum of gyroresonance emission is characterized by bright optically thick emission at low frequencies with a nearly constant brightness temperature, and a steep drop in brightness temperatures at higher frequencies. The frequency that separates the optically thick and optically thin emission corresponds to harmonics of the electron gyrofrequency and hence can be used to determine the magnetic field. On the Sun, gyroresonance emission is primarily seen from active regions when observed in the GHz frequencies. There have been a lot of studies exploring gyroresonance emission from active regions and its usage in determining the magnetic field \citep[see, e.g.,][for recent reviews]{nindos2020,gary2023}. We have detected multiple instances when the spatially resolved spectrum from a CME is consistent with having a gyroresonance origin. Under this assumed emission mechanism, we derive the magnetic field strength of the CME and its temporal evolution in the middle corona. 

In Section \ref{sec:observation}, we report the details of the observation and data analysis. In Section \ref{sec:results}, we present the results, and finally, in Section \ref{sec:discussion}, we put the results in the broader context and conclude. 

\section{Observations and data analysis} \label{sec:observation}


The data we use were recorded by the Owens Valley Radio Observatory's Long Wavelength Array (OVRO-LWA) on March 9, 2024. OVRO-LWA is an all-sky imager operating from 13.4--86.9 MHz (see \citealt{Anderson2018} for general descriptions of
the instrument). It recently completed a major expansion, which now consists of 352 antennas and has multiple solar-dedicated observing modes\footnote{An overview paper describing the recently expanded OVRO-LWA and its solar capabilities is currently under preparation.}. 
While OVRO-LWA was observing daily throughout the day, it was still under commissioning at the time. 
Hence, we have primarily used data products obtained from the standard imaging mode in this work. This mode has a time and frequency resolution of 10s and 24 kHz, respectively. Each such dataset has a data volume of $\sim 286$ MB. This translates to about 10 TB of data each day. Due to this high data volume, the visibilities are not saved for long. Instead, we use them to produce images at a one-minute cadence, which are then saved. All imaging was done using the OVRO-LWA data analysis package, details of which are provided in Mondal et al. (in preparation). While the imaging pipeline provides images averaged over a bandwidth of 0.4 and 5 MHz, the 5 MHz images, thanks to the wider bandwidth used for imaging, naturally have a higher dynamic range compared to the 0.4 MHz images. Due to the faint nature of the emission detected here, all images presented in this work have a bandwidth of 5 MHz. The time resolution of the images is 10\,s. At the low radio frequencies studied here, ionospheric refraction can shift the observed source location in the sky plane by several arcminutes. A refraction correction algorithm is implemented in the standard solar imaging pipeline of OVRO-LWA, details of which are described in Mondal et al. (in preparation). However, the algorithm can lead to spurious results in the presence of CMEs, which was the case here. Hence ionospheric shift correction was done manually, using white light images as a reference.\footnote{We have used the toolkit developed in \url{https://github.com/sageyu123/ovro-lwa-things} for this purpose.}

\begin{figure*}
    \centering
    \includegraphics[trim={2cm 0 2cm 0},clip,scale=0.4]{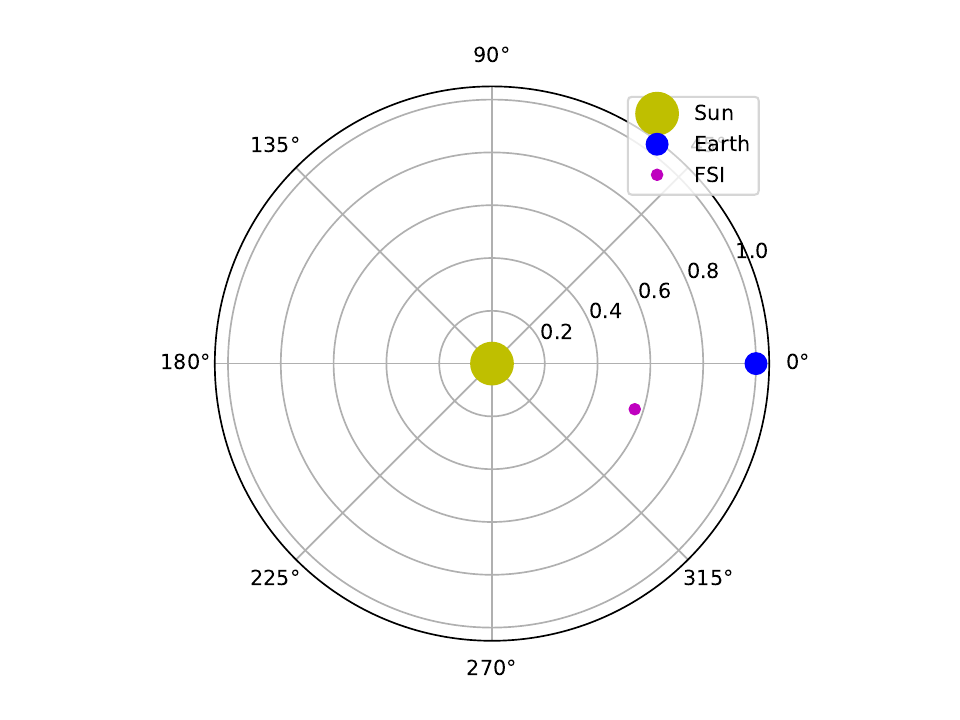}
    \includegraphics[trim={0 0 1.8cm 0},clip,scale=0.4]{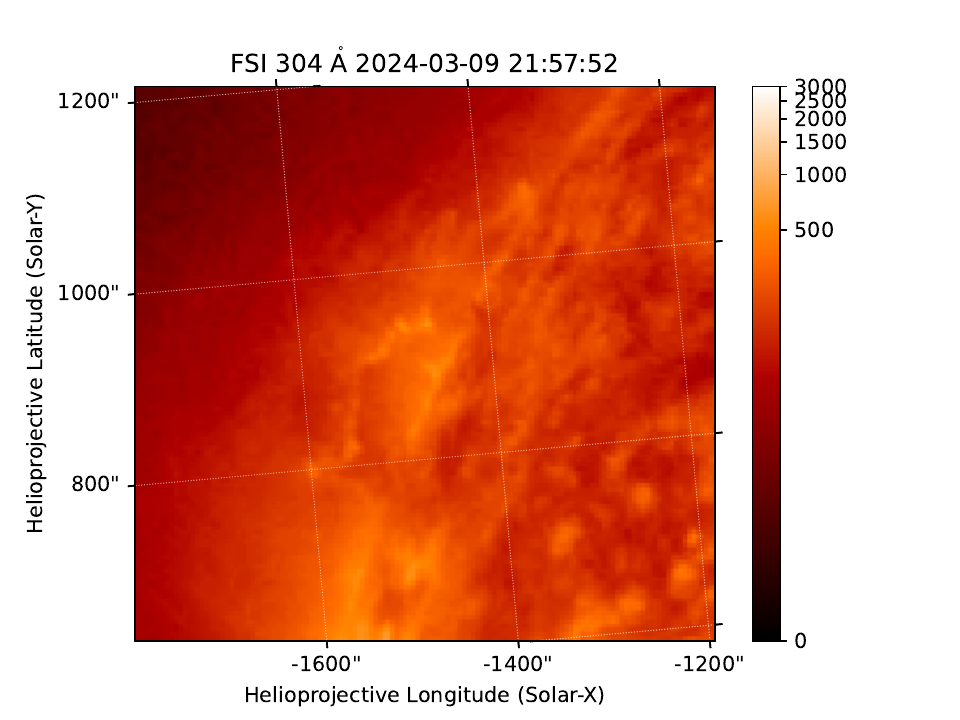}
    \includegraphics[trim={0 0 1.8cm 0},clip,scale=0.4]{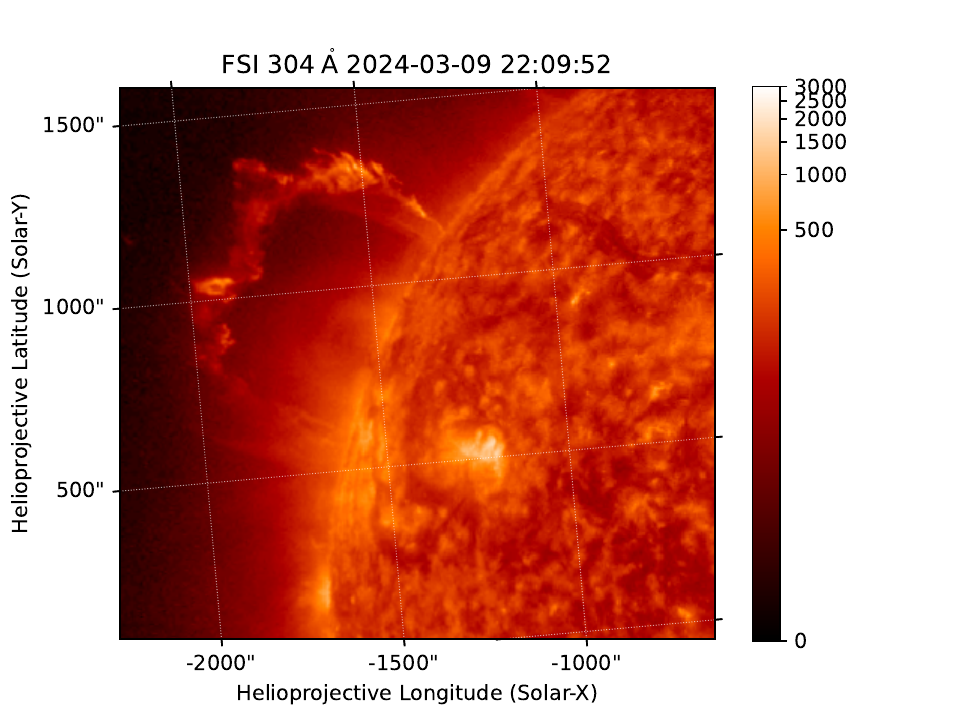}
    \caption{Left panel: Location of Solar Orbiter, Earth, and Sun. Middle and right panel: SolO/FSI 304{\AA} images at two different times showing the early times of the solar eruption. The time indicated on the title of the panels show the observation time, after correcting for the light travel time between SolO and Earth.}
    \label{fig:solar_orbiter_images}
\end{figure*}

On this day, the C2 detector of the Large Angle Spectrometric Coronagraph (LASCO) onboard the Solar and Heliospheric Observatory (SOHO) detected a white light (WL) CME at 22:24 UT.  The WL CME was associated with an eruption that originated from behind the eastern solar limb. The eruption was behind the limb even from the perspective of Solar Orbiter (SolO). SolO was trailing the Earth during this time, and had a longitudinal separation of $17.8^\circ$ (see the left panel of Figure \ref{fig:solar_orbiter_images} for the relative locations of SolO and Earth). The eruption was first detected in SolO's Full Sun Imager (FSI) extreme ultraviolet (EUV) image around 21:54, which translates to 21:57 in Earth UT time. In Figure \ref{fig:solar_orbiter_images}, we show two example images from FSI's 304 {\AA} band in the middle and right panels. 

\section{Results} \label{sec:results}

\begin{figure*}
\centering
     \includegraphics[scale=0.5]{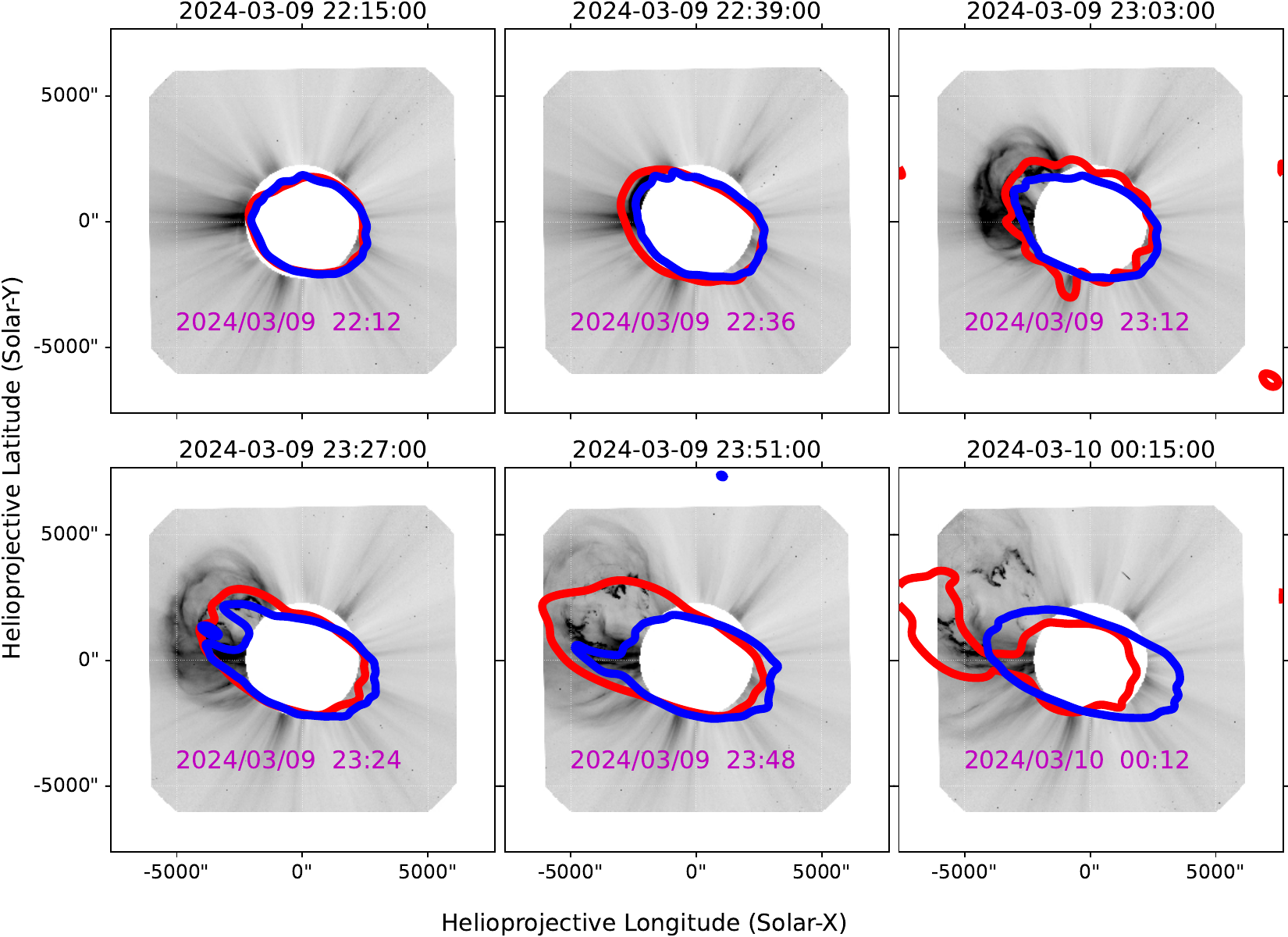}
    \caption{OVRO-LWA 39 and 80 MHz contours overlaid on LASCO C2 images at multiple timestamps beginning at a time when the CME is not yet in the LASCO C2 field of view, till the time of sunset at the observatory. The 39 and 80 MHz contours are drawn at 0.07 and 0.025 MK, respectively. The title of each panel shows the time of the radio image. The time written inside each panel shows the time of the background white-light image, which is shown in inverted grayscale (i.e., dark color is brighter). Due to the limited time cadence of LASCO C2, the nearest available white-light image is chosen as the background. } 
    \label{fig:cme_snapshots}
\end{figure*}

{ Figure \ref{fig:cme_snapshots} shows overlay radio and white light images at multiple representative times. The times corresponding to each radio image are provided in the title of each panel. The snapshots cover a large timerange, starting from an instant when the CME was not yet visible in the LASCO C2 field of view to the time, to close to sunset at the OVRO-LWA site. Later, the image quality becomes poor due to the low elevation of the sun, resulting in an elongated beam. The time corresponding to the background white-light image is also provided in each panel, using magenta color. A composite movie of the radio and white light images covering the time duration is provided in the Appendix (Figure \ref{fig:cme_snapshot_movie}). }
In Figure \ref{fig:cme_snapshots}, the red and blue contours correspond to images at 39 and 80 MHz, respectively, which are drawn at 0.07 MK and 0.025 MK, respectively.
From the radio image series, it is evident that radio emission from the WL CME is not detected prior to 22:24 UT, when it is first seen in LASCO C2 images. 
Starting from the same time, radio emission is detected from the CME in the OVRO-LWA images, till close to sunset.  We also find that the heliocentric extent of the detected radio emission is closely associated with the WL emission. The radio emission is primarily arising from features that are bright in the WL data, although not all such bright features have a corresponding radio counterpart. We have extracted the radio spectra at different regions of the CME at different times. As expected, the spectra are highly time variable and display complex structures. However, between 23:34:07 on March 9 and 00:16:05 March 10, we find multiple instances where the spectrum is consistent with it having a gyroresonance origin.

\begin{figure*}
    \centering
    \includegraphics[width=0.9\linewidth]{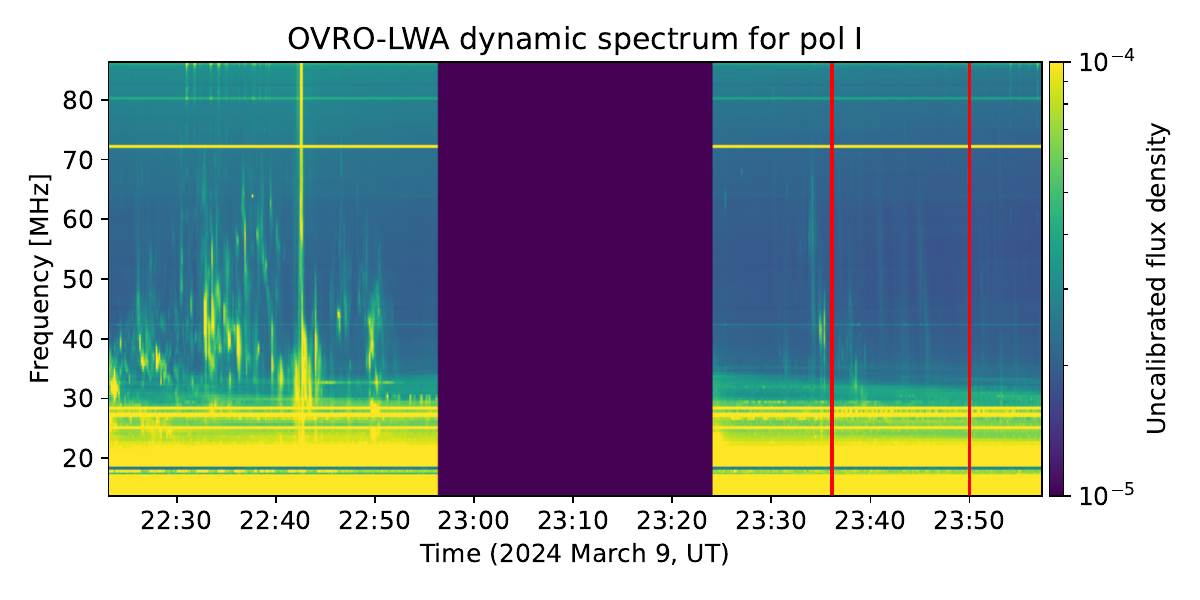}
    \caption{Shows the dynamic spectrum from OVRO-LWA in uncalibrated units. The red vertical lines show the times within the interval shown for which we find spectra consistent with gyroresonance origin. There is also a data gap between 22:56 -- 23:24 UT and has been darkened.}
    \label{fig:ds}
\end{figure*}

{In Figure \ref{fig:ds} we show the dynamic spectrum corresponding to the times presented here. While we have presented imaging data after 00 - 00:16 UT, dynamic spectrum data after the times shown were only available after 00:24 UT, and has not been shown here. As mentioned earlier, during this time the instrument was still under commissioning and there were issues with flux calibration of the beam data. Hence the dynamic spectrum data are presented in uncalibrated units. The red vertical lines show the times within the interval shown for which we find spectra consistent with gyroresonance origin.}

\begin{figure*}
\centering
\includegraphics[width=0.9\linewidth]{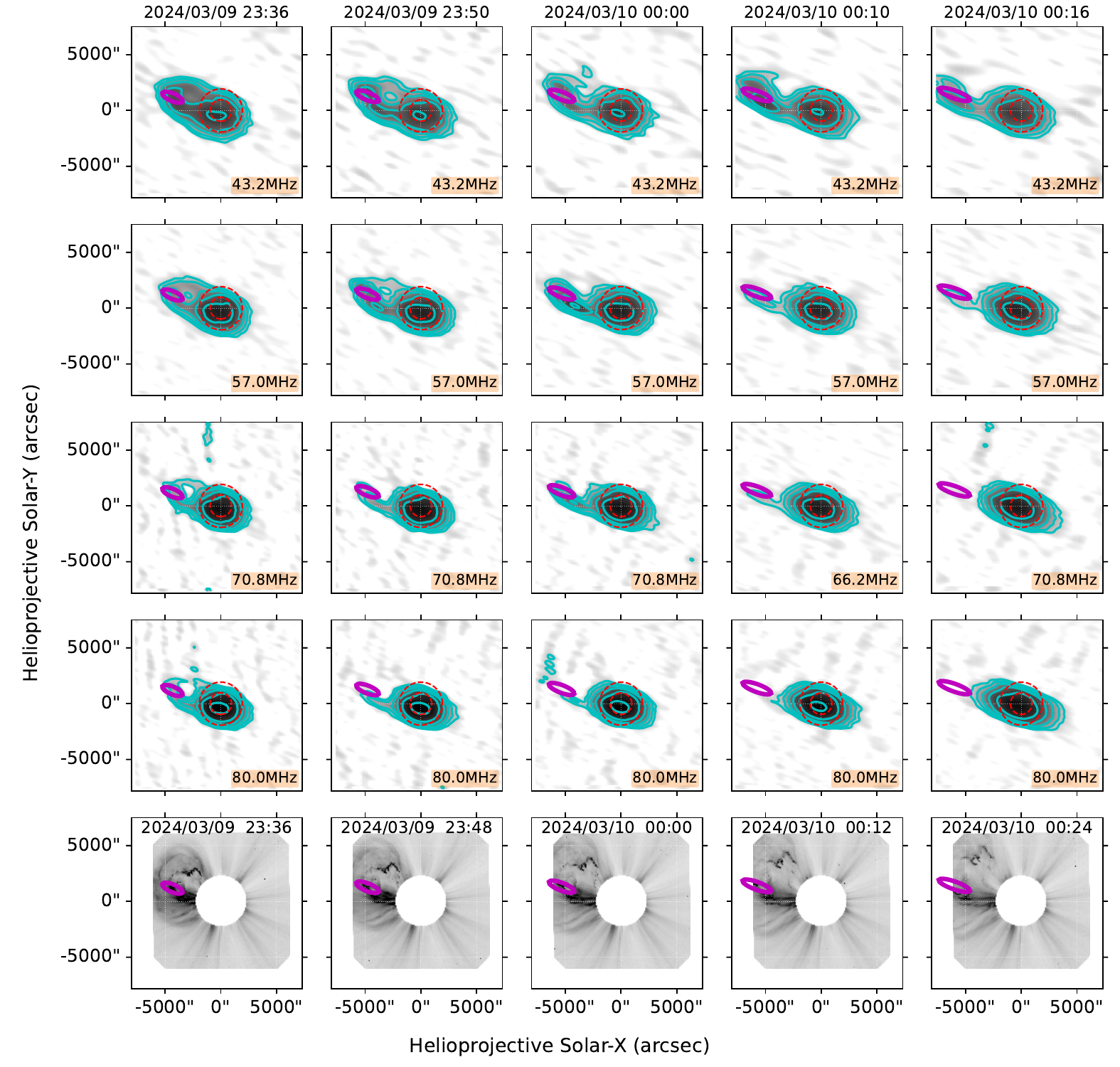}
\caption{The top four rows show OVRO-LWA radio images at a few example frequencies and times during the WL CME. The time corresponding to each column is indicated at the top of each column. The contour levels are at 0.02, 0.04, 0.08, 0.16, 0.32, 0.64, and 1 MK. In the bottom row, we have shown the nearest white light image from the LASCO C2. The time corresponding to the white light image is written in each panel. The magenta ellipse in each panel shows the region from which spectrum, shown in Figure {\ref{fig:spectra}} has been extracted.}
\label{fig:cme_multi_freq_images}
\end{figure*}

 The top four rows of Figure \ref{fig:cme_multi_freq_images} shows radio images at a few frequencies at multiple times. The overlaid contour levels at each image are at 0.02, 0.04, 0.08, 0.16, 0.32, 0.64, and 1 MK. The time corresponding to each column is indicated at the top of each column. The frequency corresponding to each image is indicated inside the respective image. The red dashed circles mark 1 and 2 $R_\odot$, respectively. In each image, we see that the radio quiet sun disk is well aligned with the disk drawn, indicating that the multi-frequency images do not have any significant residual offset after the refraction correction. Additionally, based on the brightness temperature contours, the sharp drop in brightness temperatures at the higher frequencies is also evident. 
 
\begin{figure}
    \centering
    \includegraphics[scale=0.5]{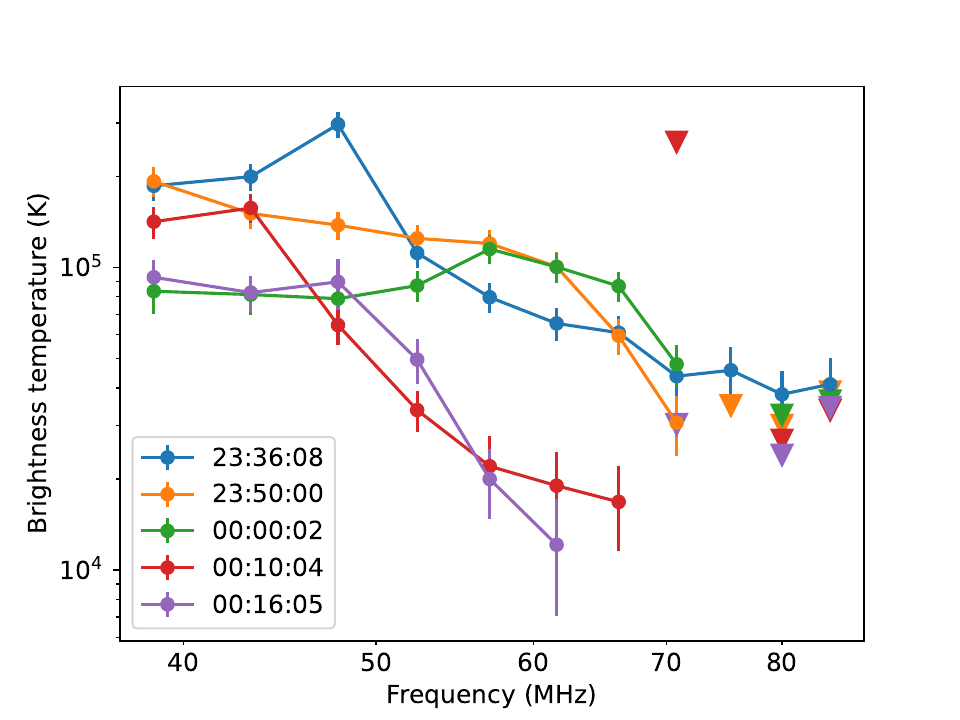}
    \caption{Spectra for different times are shown. The region from which the spectrum has been extracted is shown with an ellipse over the image at the corresponding time in Figure \ref{fig:cme_multi_freq_images}. Circles and triangles show the detections and upper limits respectively.}
    \label{fig:spectra}
\end{figure}

 This becomes clearer from the spectra shown in Figure \ref{fig:spectra}. {The spectra have been extracted from a region, which is bright in white-light (which appears dark in the image with inverted grayscale) and is located ahead of the main bulk of the CME core. The regions are indicated using magenta ellipses in Figure \ref{fig:cme_multi_freq_images}. The size of the magenta ellipse is equal to the restoring beam at 39 MHz at the corresponding timestamp. The heliocentric distance of the center of the magenta ellipse increases from 4.9 $R_\odot$ at 23:34 UT to 7.5 $R_\odot$ at 00:16:05 UT, to match the motion of the white-light feature in the sky plane. }
 The points shown with circles denote brightness temperature measurements where the peak inside the magenta region is greater than 5 times the rms of a source free region in the image. In the case that the peak is lower than this threshold, we have shown the upper limit using triangles, where the upper limits are assumed to be 5 times the rms. For each of the spectra plotted, we find that at lower frequencies, the brightness temperature is mostly constant, accompanied by a sharp drop at higher frequencies. This is consistent with the emission mechanism responsible being gyroresonance emission. {While the spectra shown here has been extracted from a particular bright white-light region, located ahead of the main bulk of the CME, the observed spectral properties are not unique to that particular white light feature. We observe similar spectra from multiple other locations which lie close to the region shown here in the sky plane. }
 
 For the gyroresonance emission spectrum, the break frequency, separating the optically thick and optically thin parts of the spectrum, corresponds to a harmonic of the electron gyrofrequency $\nu_{g}=2.8sB$~MHz, where $s$ is the harmonic number often falling between 2--4, and $B$ is the magnetic field strength in Gauss. 
The commissioning data used in this study cannot give detailed diagnostics for $s$ (but this is possible in the near future when imaging spectropolarimetry is commissioned). Assuming $s=3$ as is often the case for active regions \citep[e.g.][]{vourlidas2006, alissandrakis2021}, we estimate that the magnetic field varies between 5.6 -- 7.9 G at heliocentric distances of 4.9 -- 7.5 $R_\odot$. 

\section{Discussion} \label{sec:discussion}

Gyrosynchrotron emission and free-free emission are the only two radio emission mechanisms that have been reported from CMEs. While free-free emission can be directly ruled out based on the spectral shape presented here, gyrosynchrotron emission arising from an inhomogeneous source with a fine-tuned distribution of magnetic field/nonthermal electrons can possibly explain the observed spectra. {In the top panel of Figure \ref{fig:qualitative_fit}, we show two example model spectra from an inhomogeneous source that show similar qualitative behavior to the spectra corresponding to 00:00:02 and 00:16:05 UT shown in Figure \ref{fig:spectra}. The relevant plasma parameters used to generate the spectra are also indicated in the figures. We also used a thermal plasma temperature of 1 MK. We used the gyrosynchrotron codes implemented in \citet{fleishman2010} to compute the spectra. Apart from the magnetic field, all other parameters are assumed to be homogeneous along the line of sight (LOS). For implementing the inhomogeneity in the magnetic field, we divided the LOS into 30 voxels, and the magnetic field of each voxel is assumed to be given by $B_0/(30-i),i=0\cdots29$. { The voxel with $i=0$ is located closest from Earth along the LOS.} We note that our choice of the magnetic field variation along the line-of-sight is not based on any physical insight. Our goal is to demonstrate that a inhomogeneous magnetic field can produce the observed spectra and hence we have chosen an arbitrary, but plausible magnetic field distribution along the line-of-sight. We have denoted the thermal electron density, angle of the LOS with the magnetic field, nonthermal electron density, and depth along the LOS by $n_e$, $\theta$, $L$, respectively. We assume that the nonthermal electron distribution can be modeled as a power law between 10 keV and 10 MeV with a power-law index $\delta$. In order to match the observed brightness temperature values, we have also assumed that the gyrosynchrotron emission originates from a small region compared to the area of the ellipse from which the spectrum has been extracted. The filling factor is indicated by $\eta$, which is well below 1\% in both cases.} It should be noted that, apart from \citet{kansabanik2024}, which reported gyrosynchrotron emission from an inhomogeneous source, all other similar works were successful in modeling the Stokes I spectrum with a homogeneous source model \citep{bastian2001, maia2007, tun2013, bain2014, mondal2020, kansabanik2023, chhabra2021}. In fact, even in \citet{kansabanik2024}, the Stokes I spectrum was also well modeled by a homogeneous source model. 
The spectral shape reported in past studies was also very different from that reported here.

On the other hand, the gyroresonance emission model can explain the observed spectra with the least amount of complexity introduced in the source model (i.e., Occam's Razor principle). {Gyroresonance emission spectrum arises because of the absorption of emission of thermal electrons in the presence of magnetic field getting absorbed by the layers with a lower magnetic field strength at higher harmonics. 
In the lower panels of Figure \ref{fig:qualitative_fit}, we show two example spectra for a thermal population of electrons.

\begin{figure*}
    \centering
    \includegraphics[trim={0 0 0 0.5cm},clip,width=0.45\linewidth]{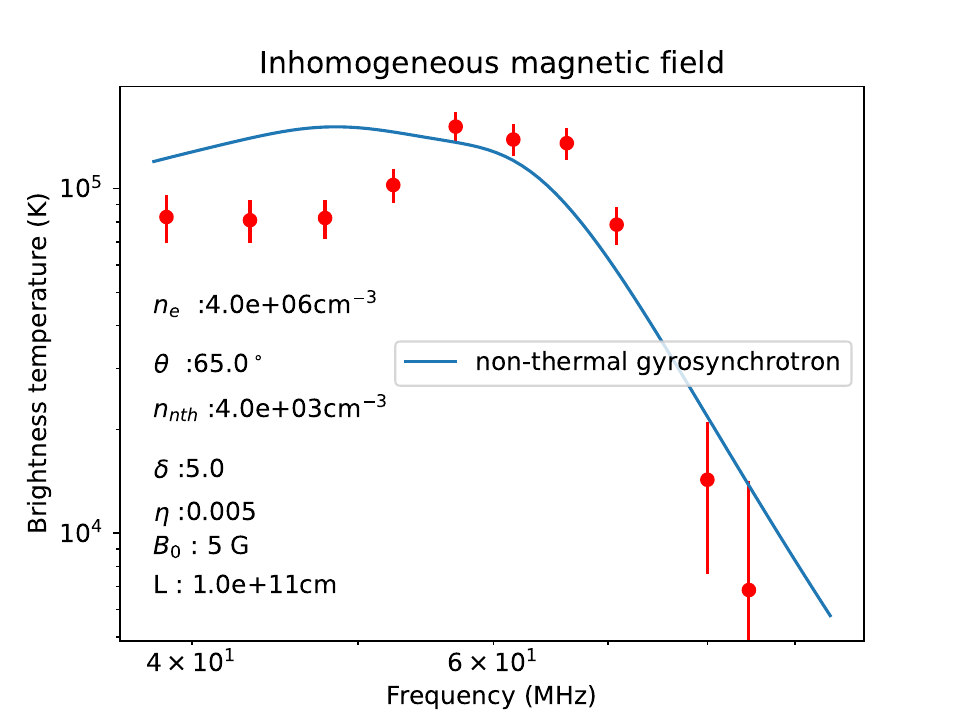}
    \includegraphics[trim={0 0 0 0.5cm},clip,width=0.45\linewidth]{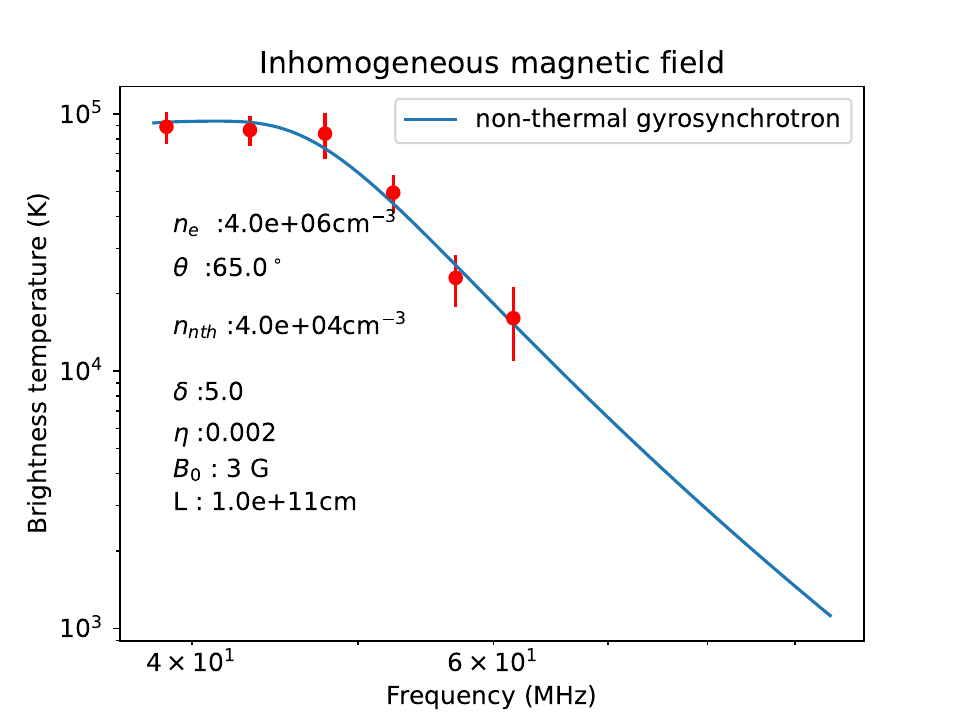}
    \includegraphics[trim={0 0 0 0.5cm},clip,width=0.45\linewidth]{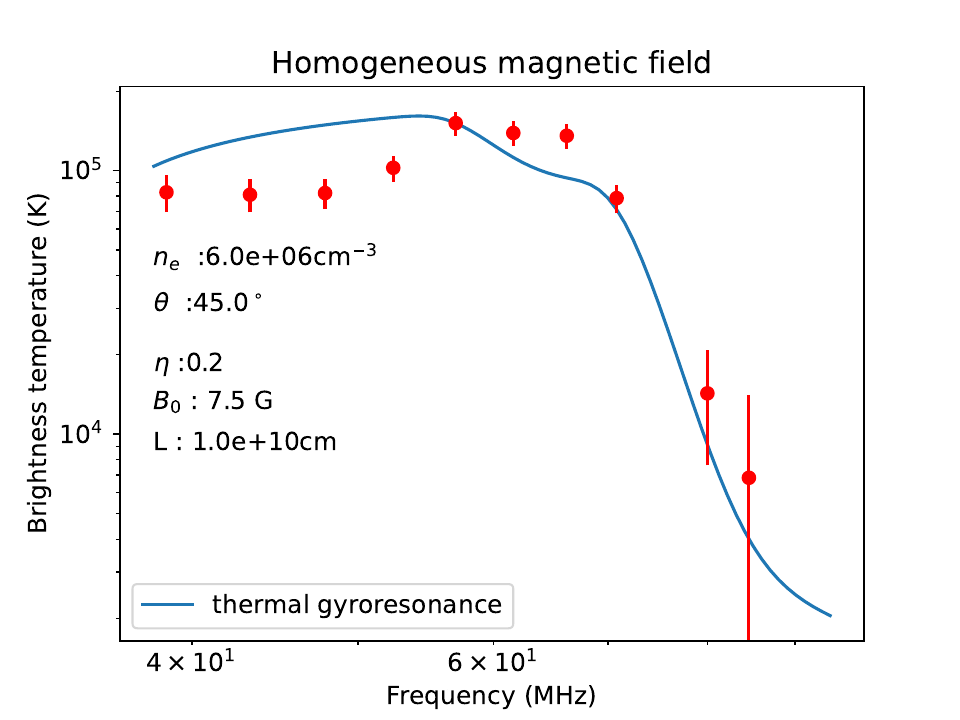}
    \includegraphics[trim={0 0 0 0.5cm},clip,width=0.45\linewidth]{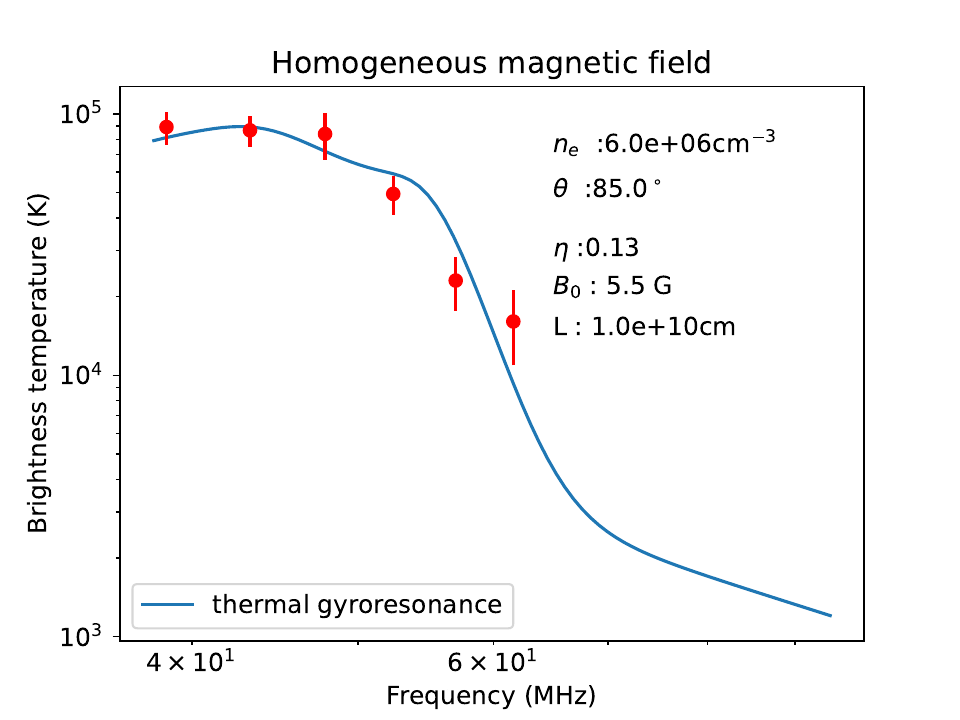}
    \caption{{ The top row shows the spectra where we have assumed that the nonthermal electron distribution responsible for the gyrosynchrotron emission follows a powerlaw distribution. The bottom row shows the thermal gyroresonance emission spectrum from a thermal distribution of electrons. The red points in the left and right columns shows the observed spectra at 00:00:02 and 00:16:05 UT (refer to Figure \ref{fig:spectra}). }}
    \label{fig:qualitative_fit}
\end{figure*}

Assuming that the break frequency is the third harmonic of the gyrofrequency, we estimated that the magnetic field in the {CME core} ranges from 7.9--5.6 G between 4.9-7.5 $R_\odot$. These numbers are very similar to the peak magnetic field values used to generate the spectra in the lower panel of Figure \ref{fig:qualitative_fit}.
To the best of our knowledge, there is no measurement of the total magnetic field entrained inside a CME at such a large height. 
\citet{bird1985} reported the line of sight (LOS) component of the magnetic field, averaged over the LOS, at these heights and it was of the order of few tens of mG. So unless the magnetic field was predominantly located perpendicular to the line-of-sight, the magnetic field measurements reported here are much larger than these early measurements. \citet{bird1985} also pointed out this apparent discrepancy between their measurements and measurements made at lower heights and suggested that this can be caused if the LOS component of the magnetic field are ordered in an incoherent manner.

{\citet{mondal2020} presented a compilation of CME magnetic field measurements at different coronal heights, determined using the gyrosynchrotron emission technique. Based on the compilation, we conclude that the CME magnetic field ranges from 1--23 G between 1--3 $R_\odot$. 
Since the height range of these earlier measurements is not the same as that of this work, we predict the value at lower heights by extrapolating the value determined here following the relationship provided in \citet{koya2024}. We find that the predicted magnetic field value at $3 R_\odot$ ranges from about 12--20 G, which is consistent with values reported earlier.  { However, it must be noted that the magnetic field dependence on the heliocentric distance mentioned in \citet{koya2024} is only from a single CME. The relationship can vary from CME to CME as well. Additionally, it is also possible that the break frequency observed in this work corresponds to a higher harmonic of the local gyrofrequency than our assumed value ($s=3$). }

In addition, for both the gyroresonance and gyrosynchrotron emission, the expected peak brightness temperature is $\gtrsim$1 MK for a source plasma temperature of coronal values (assumed to be 1 MK for the model spectra shown in Figure~\ref{fig:qualitative_fit}). However, the observed peak brightness temperature is often smaller by an order of magnitude, around 0.1\,MK. Similar to \citet{mondal2020} and \citet{kansabanik2023}, we also hypothesize that this is occurring because the magnetic field, responsible for producing the emission, is present only in a small region compared to the area covered by the restoring beam.  In Figure \ref{fig:qualitative_fit}, we also see that the filling factor $\eta$ can be quite small for the model spectra to have a similar level of peak brightness temperature. For the thermal distribution, we have used a filling fraction of $20\%$, which is also comparable to earlier works. We have used source sizes of about 0.03 $R_\odot$ and 0.25 $R_\odot$ for the nonthermal and thermal electron distributions respectively.  One thing to note here is that the GS spectrum modelling code ignores the gyrosynchrotron contribution from the thermal electron population, unless specifically a thermal or a combination of thermal and nonthermal electron population is chosen. In reality, the contribution from nonthermal electrons is always on top of the contribution from thermal electrons. Hence, physically the predicted GS emission shown in the top panel of Figure \ref{fig:qualitative_fit} is only a lower limit. Considering that the emission area is a multiplicative term when computing the flux density, this implies that the emission area used here can be considered as an upper limit, given all other parameters remain same.

{The assumption that the emission is primarily arising from small knots of magnetic field can also explain the absence of smooth variation of magnetic field with heliocentric distances. The smooth variation of magnetic field is expected due to global expansion of the CME. However small-scale magnetic structures, if produced due to evolution of the CME as well or its interaction with the surrounding heliospheric structures, need not follow this expectation. The magnetic field of the structures produced in this manner may depend on details of the generation process and hence can show very different variation with heliocentric distance.}

{High resolution imaging observations available with the Widefield Imager for Parker Solar Probe \citep[WISPR;][]{vourlidas2016} have also revealed the presence of multiple small-scale magnetic structures inside a CME \citep{cappello2024}. Visually we estimate that the length scales of the structures observed in \citet{cappello2024} ranges from about 120--1000 Mm, which is similar to lengthscale of the emitting region required to explain the observed brightness temperatures.} Simulations  also show considerable fine-scale structure in CMEs when they are still in the lower corona \citep{karpen2012, lynch2025}. These small structures can travel into the interplanetary medium with the CME and become visible in the WISPR data \citep{lynch2025}. The spatio-temporal evolution of these small-scale magnetic features can be quite different from the overall evolution of the CME, and hence the ``adiabatic expansion" assumption used to often describe the CME, may not be valid for these small-scale features.}

To summarize, this work presents multiple example spectra from a CME, which are consistent with the emission having a gyroresonance origin. We have presented the spectrum at multiple times and have shown that, as expected, the break frequency decreases as time progresses and the CME moves to higher heights. With the improvement in our understanding of this emission from CMEs and the consequent identification of the most probable harmonic of the break frequency, it has the potential to serve as another method to probe the CME magnetic field.

\begin{acknowledgments}
S.M., B.C., and D.G. were supported by the NASA Living With a Star (LWS) Science grant 80NSSC24K1116. The OVRO-LWA expansion project was supported by NSF under grant AST-1828784. OVRO-LWA operations for solar and space weather sciences are supported by NSF under grant AGS-2436999. P. Z. acknowledges support for this research by the NASA Living with a Star Jack Eddy Postdoctoral Fellowship Program, administered by UCAR’s Cooperative Programs for the Advancement of Earth System Science (CPAESS) under award 80NSSC22M0097.
\end{acknowledgments}

\appendix

In Figure \ref{fig:cme_snapshot_movie} we show a composite movie of radio and white light images (movie is available as a supplementary material). In each frame of the movie, we have overlaid radio contours at 39 and 80 MHz on top of the nearest available white-light image. The 39 and 80 MHz contours are drawn at 0.07 and 0.025 MK respectively. The title of each panel shows the time of the radio image. The time written inside each panel shows the time of the background white-light image. Due to dynamic range limitations, some frames lack contours at one or both the frequencies. In the left panel, we show the images after refraction correction, and correspond to the images which have been used in this work. In the right panel, we have shown the original images before any refraction correction were performed.

\begin{figure}
    \begin{interactive}{animation}{cme_video.mp4}
    \includegraphics[scale=0.5]{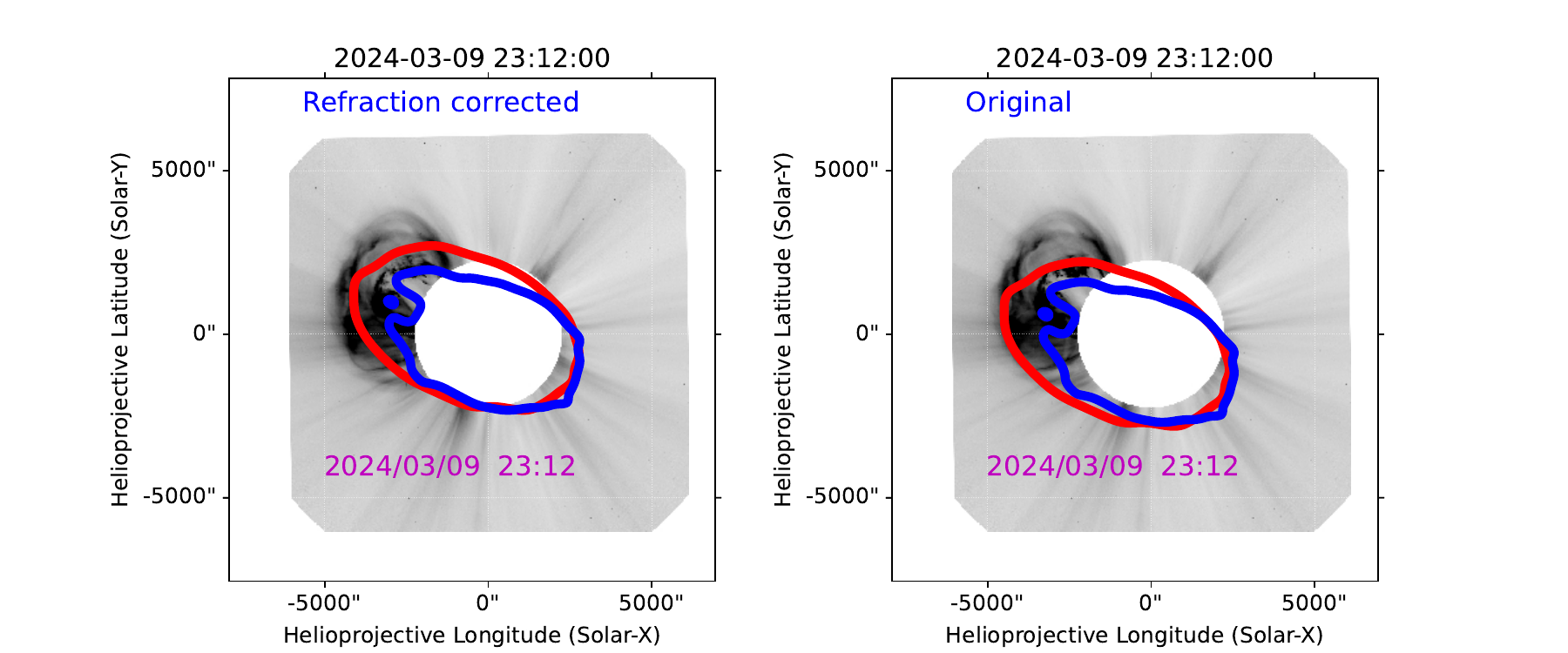}
    \end{interactive}
    \caption{Shows overlay of 39 and 80 MHz contours at an example timeslice. The 39 and 80 MHz contours are drawn at 0.07 and 0.025 MK respectively. The title of each panel shows the time of the radio image. The time written inside each panel shows the time of the background white-light image. The accompanying animation (available online only) shows these overlays at a cadence of 1 minute, beginning at a time when the CME is not yet in the LASCO C2 field of view, till the time of sunset at the observatory. Due to the limited time cadence of LASCO C2, the nearest available while-light image is chosen as the background. Radio images at a few time-frequency slices are dynamic-range limited, and hence contours corresponding to those have not been shown. Right and left panels show the images before and after refraction correction.} 
    \label{fig:cme_snapshot_movie}
\end{figure}
\bibliography{bibliography}
\bibliographystyle{aasjournal}



\end{document}